\documentclass{emulateapj}

\slugcomment{Revision of ApJ 647, L99 [2006] for Erratum}

\shorttitle{Mass drives Spheroidal Galaxies Evolution}
\shortauthors{di Serego Alighieri et al.}

\begin{document}


\title{The Influence of Mass and Environment on the Evolution of Early-Type Galaxies}


\author{Sperello di Serego Alighieri} 
\affil{INAF -- Osservatorio Astrofisico di Arcetri, Largo E. Fermi 5,
50122 Firenze, Italy}
\email{sperello@arcetri.astro.it}

\author{Barbara Lanzoni}
\affil{INAF -- Osservatorio Astronomico di Bologna, Via Ranzani 1, 
40127 Bologna, Italy}

\and

\author{Inger J\o rgensen}
\affil{Gemini Observatory, 670 North A'ohoku Place, Hilo, HI 96720}

%

\begin{abstract}
We report on a uniform comparative analysis of the fundamental parameters of
early--type galaxies at $z$$\sim$1 down to a well defined magnitude limit 
($M_B\leq -20.0$ in the field and $M_B\leq -20.5$
in the clusters). The changes in the ${\cal M}/L_B$ ratio from $z$$\sim$1 to
today are larger for lower mass galaxies in all environments, and are similar
in the field and in the clusters for galaxies with the same mass.
By deriving ages from the ${\cal M}/L_B$ ratio, we estimate the formation 
redshift for early-type galaxies as a function of galaxy mass and environment. 
We find that the age of early-type galaxies increases with galaxy
mass (downsizing) in all environments, and that cluster galaxies appear to have 
the same age within 5\% as field galaxies at any given galaxy mass.
The first result confirms similar ones obtained by other
means, while the second one is controversial. The most recent incarnation of the hierarchical
models of galaxy formation and evolution is capable of
explaining the first result, but predicts that cluster galaxies should
be older than field galaxies.
We also find a total lack of massive early--type galaxies
(${\cal M}>3\times 10^{11} {\cal M}_{\sun}$) with a formation redshift
smaller than 2, which cannot be due to selection effects.
\end{abstract}


\section{Introduction}

Early--type galaxies (ETG) contain most of the visible mass in the Universe
\citep{ren06}
and are thought to reside in the highest density peaks of the underlying dark 
matter distribution. Therefore,
understanding their evolution is crucial for understanding the evolution of 
galaxies and structures in general. 
In the 3-dimensional space of their
main parameters (the effective radius $R_e$, the central velocity dispersion
$\sigma$, and the average surface luminosity within $R_e$,  $\langle I\rangle_e 
= L/2\pi R_e^2$), ETG concentrate on a plane thus called the Fundamental
Plane (FP, Djorgowski \& Davis 1987; Dressler et al. 1987). This
implies that, besides being in virial equilibrium, ETG show a striking
regularity in their structures and stellar populations (e.g.,
Renzini \& Ciotti 1993), which allows, at least at a first order, to use
their main observables for deriving the galaxy mass and ${\cal M}/L$ ratio.
For instance, assuming $R^{1/4}$ homology, the mass is given by (Michard 1980;
see also Cappellari et al. 2005):

\begin{equation}
{\cal M} = 5R_e\sigma^2/G.
\end{equation}

Moreover, the slope of the FP can be interpreted as a systematic variation of
the ${\cal M}/L$ ratio along the plane by a factor of $\sim 3$ (e.g., Ciotti,
Lanzoni \& Renzini 1996). At high redshift the FP is known to
stay thin, and its intercept shows an offset with respect to the local one
that corresponds to a change in ${\cal M}/L$ consistent with passive luminosity
evolution (see Renzini 2006 and references therein). If ascribed to
differences in the stellar populations, the observed changes in
the ${\cal M}/L$ ratio can be used to infer the ages of ETG. We report on a
comparative analysis of the best data on the fundamental parameters of ETG at
$z$$\sim$1, the highest redshift for which these data are currently available,
obtained from recent spectroscopic observations with 8-10m class telescopes, 
complemented with deep imaging with the Hubble Space Telescope. Using
the Universe as a time machine and profiting from the large leverage provided 
by the redshift, we infer ages for ETG and analyse them as a function of
galaxy mass and environment. We assume a flat Universe with $\Omega_m=0.3$,
$\Omega_{\Lambda}=0.7$, and $H_0=70$ \mbox{km} s$^{-1} {\rm Mpc}^{-1}$, and
we use magnitudes based on the Vega system.

\section{Background}

Recent studies \citep{dsa05,tre05,vdw05} of the FP of ETG in the field at $z$$\sim$1, in 
the rest-frame B-band, down to relatively faint luminosities ($M_B\leq-20.0$), and 
hence {small} masses, demonstrate that, in addition to the offset, the FP at 
$z$$\sim$1 also shows a different slope. This implies that the galaxy 
${\cal M}/L_B$ ratio evolves with redshift in a way that depends on the galaxy mass.
By comparing the ${\cal M}/L_B$ ratio of 
field ETG to that of massive (${\cal M}>10^{11}{\cal M}_{\sun}$) ETG in clusters,
a faster evolution of ${\cal M}/L_B$ for the less massive galaxies has been 
derived, and it is interpreted as a manifestation of 
downsizing, i.e. the tendency of smaller galaxies to have later or more prolonged 
star formation histories than the massive ones \citep{cow96}.

Very recently the high-z FP of the ETG has been studied in two clusters 
(RX J0152.7$-$1357 at z=0.835 and RX J1226.9$-$3332 at z=0.892), reaching a
similarly faint limiting  absolute magnitude ($M_B\leq-20.5$), also in the 
rest-frame B-band \citep{jor06}.
This has pointed out that, also in the clusters, the slope of the FP changes
with redshift, a manifestation of downsizing even in high density
environments. 

Unfortunately, because of an error in the calibration of the galaxy
luminosities used by \citet{jor06}, the photometry for the two clusters
should be offset to brighter luminosities
with a factor $(1+z)$. Correcting for this error corresponds to an
offset in $log L$ to brighter luminosities with $log (1+z)$, which
is 0.26 and 0.28 for RX J0152.7--1357 ($z=0.835$) and RX J1226.9+3332
($z=0.892$)
respectively. Therefore the cluster data, which we have used in the published
version of this letter (ApJ 647, L99), should be changed and we present here
a corrected version of our original letter (see also the Erratum to ApJ 647,
L99).

\section{The formation epoch of cluster and field early--type galaxies}

We make a uniform comparison of these results on the high redshift FP 
(Fig. 1) and on the consequent variations of the ${\cal M}/L_B$ ratio (Fig.
2), both in the field, by using the samples of \citet{dsa05} and of \citet{tre05},
and in the clusters, by using the sample of \citet{jor06}. 
As a reference in the local Universe, we use new data for 
the Coma cluster \citep{jor99,jor06}. The figures show that the change in 
${\cal M}/L_B$ between ETG at high redshift and the local ones decreases with
the galaxy mass and
is very similar in the clusters and in the field. However, since the clusters are 
at a slightly lower redshift, a deeper analysis is 
necessary to show this more clearly. The usual way to achieve this purpose is to compare the ${\cal M}/L_B$ ratio
of the high redshift ETG with the corresponding ratio obtained for massive (${\cal M}\geq 
10^{11}{\cal M}_{\sun}$) cluster ETG at the same redshift, as compiled and parameterized by 
\citet{van03}. However this analysis is unsatisfactory, 
since the massive cluster ETG are not necessarily a uniform class, and, by
construction, such a procedure prevents one from studying
the lower mass cluster galaxies. What is of interest is how the star formation history of ETG, 
or at least their average stellar age, de\-pends on both galaxy mass and environment. 
We analyse this by interpreting the changes in ${\cal M}/L_B$ as 
differences in the ages of the stellar populations\footnote{It has been shown that other possible interpretations, i.e. systematic structural 
changes and partial support by rotation, can only explain a small fraction of the 
observed differential evolution of ${\cal M}/L_B$, and that this evolution 
correlates with the rest-frame $U-B$ colour, thereby providing 
independent evidence for changes in the stellar populations \citep{dsa05}.}. 
While the star formation
histories of some ETG could have had multiple episodes of star formation
\citep{tre05}, we can only estimate luminosity weighted average stellar ages,
by using single stellar population models.
We therefore infer galaxy ages using the relation between 
${\cal M}/L_B$ and age obtained by evolutionary population synthesis models
(Maraston 2005, see also
\rm{http://www-astro.physics.ox.ac.uk/$\sim$maraston/SSPn/ml/}), and assuming 
that the model stellar mass is proportional to the dynamical mass obtained
from equation (1).
Since the ${\cal M}/L$--age relation depends both on the
stellar initial mass function (IMF) and on the metallicity, we adopt a
\citet{kro01} IMF, which is known to better reproduce the characteristics of
low and high redshift galaxies. Moreover we estimate the galaxy metallicity
from the observed velocity dispersion following Thomas et al.  (2005; see
also Annibali et al. 2006), and we assume that this relationship does do not evolve with
redshift, as in the case of passive evolution. Given the values of ${\cal
M}/L_B$ and metallicity for every galaxy in our sample, ages have been inferred
by means of a spline interpolation of the population synthesis model results.
Then, the lookback time to formation has been
derived by using the Universe as a time machine and exploiting the large leverage 
provided by the considerable distance of the observed ETG.
The uncertainties on the age
estimates have been computed taking into account the known errors in ${\cal
M}/L_B$, as well as the uncertainties on the estimated metallicities,
due to the known errors in the velocity dispersion measurements
and to the observed scatter in the metallicity vs. velocity dispersion relation
\citep{tho05}.

\begin{figure}
\epsscale{1.70}
\plotone{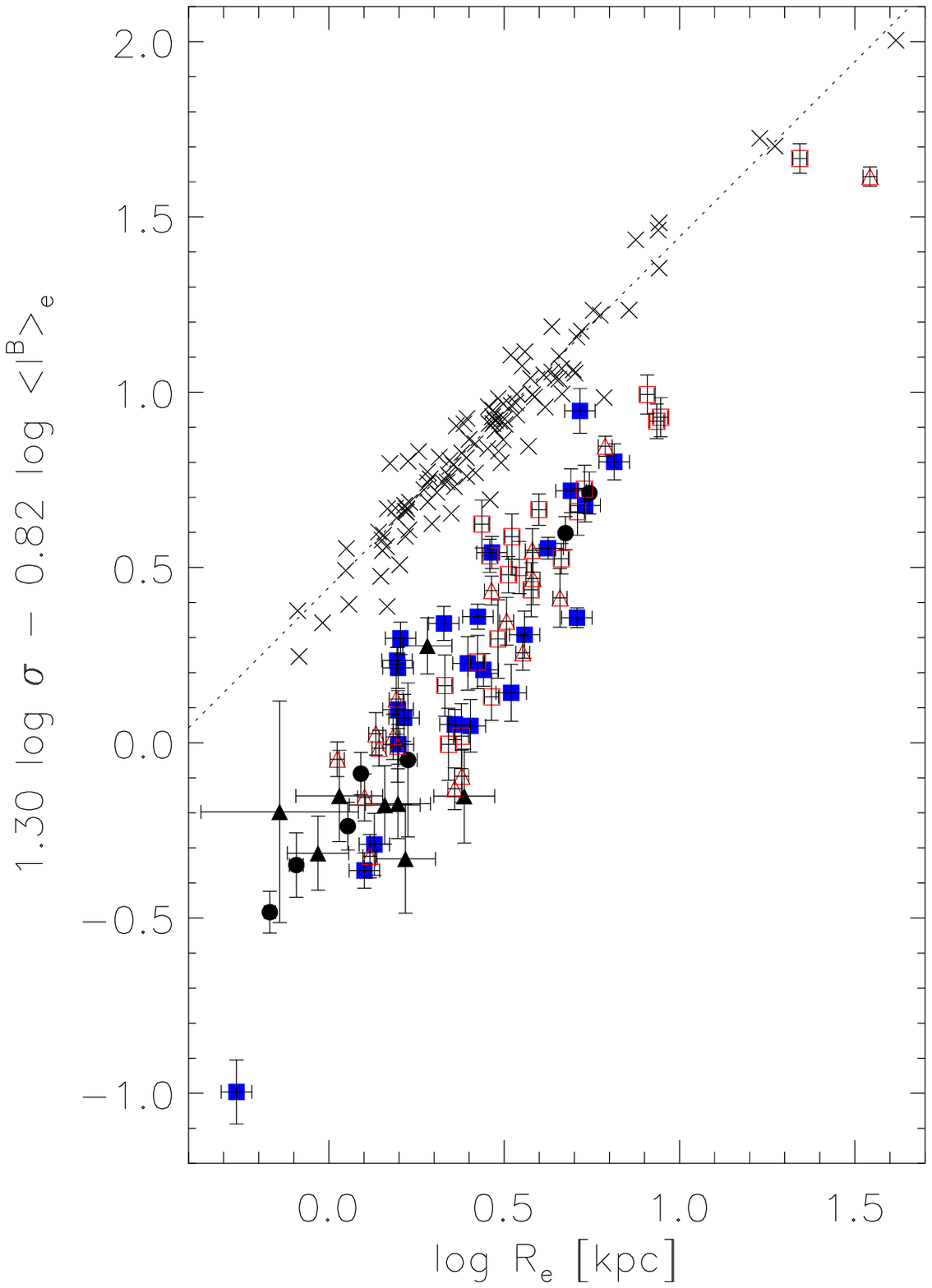}
\caption{The Fundamental Plane seen edge--on for local ETG in the Coma
Cluster \citep{jor06} (black crosses), for field ETG at $z$$\sim$1 from the
K20 survey \citep{dsa05} both in the CDFS field (filled black circles)
and in the Q0055 field (filled black triangles), for field ETG at $z$$\sim$1
in the GOODS area \citep{tre05} (filled blue squares), and for the ETG in two
clusters \citep{jor06} at z=0.835 (open red squares) and at z=0.892 (open red
triangles). The dashed line is the best fit plane to the Coma cluster
galaxies.
Compared to the local one, the FP at high redshift is offset and rotated in
all environments.}
\end{figure}

\begin{figure}
\epsscale{1.2}
\plotone{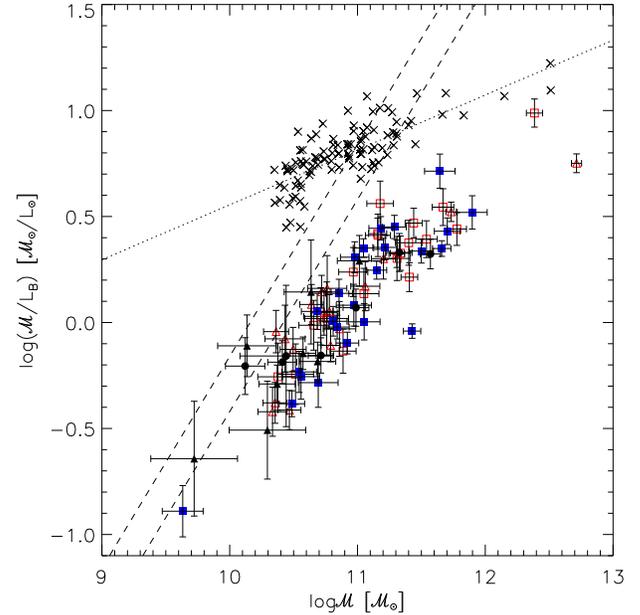}
\caption{The ${\cal M}/L$ ratio in the B-band as a function of galaxy mass
for the ETG samples shown in figure 1 (same symbols). The dotted line is
a fit to the Coma ETG, while the upper and lower dashed lines represent the
$M_B=-20.0$ and
$M_B=-20.5$
magnitude limits of di Serego Alighieri et al. (2005) and of J\o rgensen et
al. (2006) respectively. The changes in ${\cal M}/L_B$ from
high redshift to $z=0$ decrease with galaxy mass in all environments and are
similar in the field and in the clusters.}
\end{figure}

The resulting formation epochs of ETG are shown in Figure 3 as a
function of galaxy mass, both for the cluster and for the field environment.  The
estimated ages for the {two}
brightest cluster galaxies (\#1567 in RXJ0152.7-1357, and \#563 in
RXJ1226.9+3332 \citet{jor06}), are $23.4\pm 2.6$ and $16.4\pm 0.9$ Gyr,
respectively, and are not included in Fig. 3 (see below for a discussion about 
these large ages). A clear and important result is the lack of young
massive ETG. In particular all ETG with ${\cal M}>3\times 10^{11} 
{\cal M}_{\sun}$ have a lookback time to formation larger than 10
Gyr and have a formation redshift larger than 2. Clearly this
cannot be the result of a selection effect, since relatively young massive ETG
could not escape from the available surveys.
	  
\begin{figure}
\epsscale{1.15}
\plotone{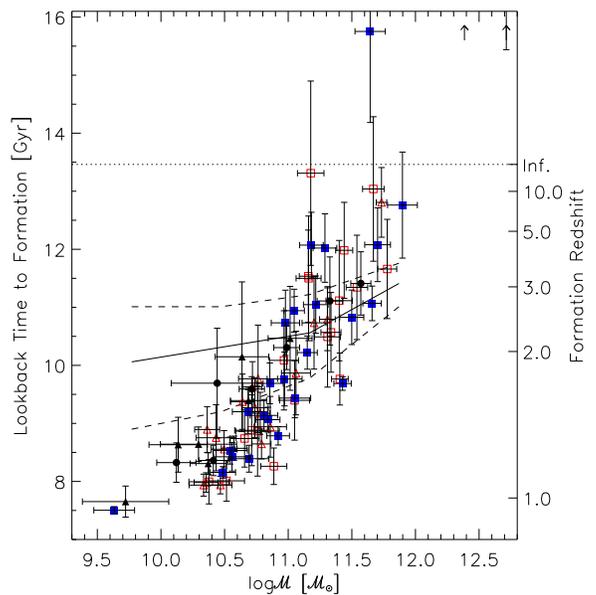}
\caption{The formation epoch for the ETG shown in figure 1
(same symbols), evaluated as explained in di Serego Alighieri et al. (2006). The
two upward pointing
arrows indicate that the two most massive cluster ETG are out of the figure
(their ages amount to 16.4 and 23.4 Gyr). The continuous line
shows the median model ages obtained by \citet{del06} from a semianalytic
model of hierarchical galaxy evolution, while the dashed lines are their
upper and lower quartiles.
More massive ETG form earlier in all environments, and the ages are not
influenced by the environment.}
\end{figure}

Confirming the analysis of the evolution of ${\cal M}/L_B$ given at the
beginning of this section, we find that more 
massive galaxies are older than lower mass ones in all environments, and that cluster 
galaxies have the same age within 5\% as field galaxies with the same mass, in the whole mass range (see
Fig. 4). 
A similar dependence of the age on the mass has 
already been obtained by an analysis of the 
absorption line indices of a sample of local ETG \citep{tho05}. However Thomas
et al. (2005) find that ETG in clusters are older than those in the field by
about 2 Gyr. Given the number of objects in the samples that we have examined
and the errors in the estimate of their age, we should have seen such a systematic 
age difference, if it were present in the data that we have used.
We argue that using the Universe as a time machine  should be more powerful than 
``archaeology'' on local galaxies, since galaxies are caught closer to the
action. Interestingly, also the Coma ETG show the
downsizing effect, and their formation redshifts are very consistent with those of
$z$$\sim$1 ETG \citep{jor99}. This suggests that the
$z$$\sim$1 samples examined here are not much affected by the progenitor bias
\citep{van01,dsa05}.
Thus, our results suggest that the first
ETG to form are the most massive ones independently of the environment.

\begin{figure}
\epsscale{1.2}
\plotone{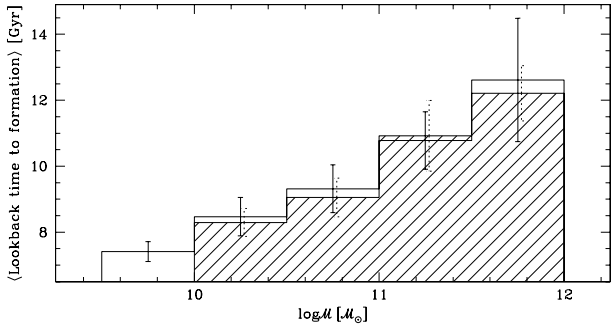}
\caption{Histogram of the average lookback time to formation per mass bin for
the high
redshift ETG in the field and in the clusters (hatched). The error bars
(dotted for the clusters) show the standard deviation due to the
galaxy--to--galaxy variations in each mass bin.}
\end{figure}

Although the absolute ages that we derive are somewhat model dependent, 
are affected by an approximate metallicity estimate, and obviously depend also 
on the adopted cosmological parameters, we stress that the {\it trends of age 
differences} between high redshift and local ETG, and between galaxies with 
different masses and in different environments are much more robust.

One of the uncertainties affecting the age estimates derives from the
assumption of structural homology when computing masses through equation (1).
It is well known that ETG show a systematic departure from
homology, both locally \citep{cao93,gut04,gav05} and at $z$$\sim$1 \citep{dsa05} and
that more precise dynamical masses can be obtained taking these deviations
into account, using the S\'ersic (1968) profile, 
to describe the observed surface brightness distribution, instead of the
$R^{1/4}$ law \citep{ber02}. These
mass estimates can be up to $\sim 50\%$ higher than those
obtained assuming homology for the low mass galaxies, but can also be
lower by up to 20\% for the high mass galaxies \citep{dsa05}. Unfortunately we do not have
S\'ersic indices for all the ETG examined here, but we have checked on the K20
field samples of \citet{dsa05} that the ages estimated by taking
non homology into account do not vary substantially from
those given in Fig. 3 and 4, computed using eq. (1).
Since the brightest cluster galaxies are known to deviate from the $R^{1/4}$
profile, and if the influence of dark matter increases in high mass galaxies,
these factors could lead to an overestimate of the the ages of the most
massive ETG in the cluster sample.

The influence of selection effects is shown by the dashed lines in
Fig. 2, which represent the magnitude limit of the K20
field samples of \citet{dsa05} and of the two high redshift clusters of 
\citet{jor06}. These samples are affected by selection only for 
${\cal M}<4\times 10^{10}{\cal M}_{\sun}$, while the different slope in the high
redshift samples compared to the local one is clearly visible also for larger
masses, thus cannot be totally due to selection effects (see also \citet{vdw05}).

Very recently the largest high resolution simulation of the growth of cosmic 
structure in the hierarchical formation scenario (the Millennium Run, Springel 
et al. 2005) has been used to study how the ages of ETG depend on environment and 
on galaxy mass \citep{del06}. In this model, since merging of smaller galaxies 
is an important ingredient for the formation of ETG, the galaxy formation time, 
which is when most of its stars formed, and the galaxy assembly time, which is 
when stars assembled in the single galaxy that we observe, are considered 
separately \citep{del06}. Our dating based on changes in the ${\cal M}/L_B$ ratio 
relates to when the stars formed, rather than to when they assembled. 
The semi-analytic hierarchical model of \citet{del06} is able to reproduce the 
already known result, i.e. that the formation times are earlier for more 
massive ETG, although the downsizing effect is considerably steeper in
the model than in the data (see
Fig. 3), but clearly predicts that cluster galaxies should be older than
field galaxies, which is not what we observe.

\acknowledgments

We thank Sandro Bressan, Claudia Maraston and Alvio Renzini for useful advice
and suggestions.
The data on high redshift field ETG were obtained at the European Southern
Observatory, Chile (ESO Programme 70.A-0548) and at the W.M. Keck Observatory 
on Mauna Kea, Hawaii.
The spectroscopic data for the high redshift clusters were
obtained at the Gemini Observatory (GN-2002B-Q-29, GN-2004A-Q-45),
which is operated by AURA, Inc., under a cooperative
agreement with NSF on behalf of the Gemini partnership: NSF
(US), PPARC (UK), NRC (Canada), CONICYT (Chile),
ARC (Australia), CNPq (Brazil) and CONICET (Argentina).
Based on observations made with the NASA/ESA Hubble Space Telescope. 


\end{document}